%% file: 00_main.tex
\documentclass[conference]{IEEEtran}
\def\IEEEtitletopspace{18pt}
\pdfoutput=1 

\usepackage[top=54pt, left=54pt, right=54pt, bottom=54pt]{geometry}
\IEEEoverridecommandlockouts

\usepackage{amsmath,amssymb,amsfonts}
\usepackage{algorithmic}
\usepackage{multicol}
\usepackage{graphicx}
\usepackage{hyperref}
\usepackage[numbers]{natbib}
\usepackage{textcomp}
\usepackage{xcolor}
\hypersetup{bookmarksopen,bookmarksnumbered,
pdfpagemode=UseOutlines,
colorlinks=true,
urlcolor=black,
linkcolor=red,
anchorcolor=red,
citecolor=red,
filecolor=red,
menucolor=red
}
\usepackage{cleveref}
\usepackage{booktabs}
\usepackage{tikz}
\usepackage{siunitx}
\usepackage[font=footnotesize]{caption}
\usepackage{subcaption}
\usepackage{nicefrac}
\usepackage{soul}
\usepackage{multirow}
\usepackage{supertabular}

\def\BibTeX{{\rm B\kern-.05em{\sc i\kern-.025em b}\kern-.08em
    T\kern-.1667em\lower.7ex\hbox{E}\kern-.125emX}}
\begin{document}

\title{\LARGE \bf Traffic Simulations: Multi-City Calibration of Metropolitan Highway Networks}

\makeatletter
\newcommand{\linebreakand}{%
  \end{@IEEEauthorhalign}
  \hfill\mbox{}\par
  \mbox{}\hfill\begin{@IEEEauthorhalign}
}
\makeatother

\author{Chao Zhang$^{1,2}$, Yechen Li$^{2}$, Neha Arora$^{2}$,\\ Damien Pierce$^{2}$, and Carolina Osorio$^{2,3}$
\thanks{$^{1}$Corresponding author, chaozhng@google.com}%
\thanks{$^{2}$Google Research}%
\thanks{$^{3}$HEC Montr\'eal}%
}

\maketitle

\begin{abstract}
This paper proposes an approach to perform travel demand calibration for high-resolution stochastic traffic simulators. It employs abundant travel times at the path-level, departing from the standard practice of resorting to scarce segment-level sensor counts. The proposed approach is shown to tackle high-dimensional instances in a sample-efficient way. For the first time, case studies on 6 metropolitan highway networks are carried out, considering a total of 54 calibration scenarios. This is the first work to show the ability of a calibration algorithm to systematically scale across networks. Compared to the state-of-the-art simultaneous perturbation stochastic approximation (SPSA) algorithm, the proposed approach enhances fit to field data by an average 43.5\% with a maximum improvement of 80.0\%, and does so within fewer simulation calls.
\end{abstract}


\section{Introduction}
\label{sec:intro}
Metropolitan transportation systems are increasingly complex due to increased traveler and vehicle connectivity, on-demand responsiveness of services, as well as increased offerings of multi-modal and shared mobility services. Cities have also witnessed an increase in the number of mobility services offered by private sector stakeholders. 
At the metropolitan scale, the higher number of both services offered and stakeholders involved makes it increasingly difficult to evaluate the performance of novel services, let alone to optimally operate and price these services. 
To address this challenge, cities are developing digital twins of their mobility systems. These digital twins come in the form of high-resolution traffic simulators that provide a detailed description of both (i) the services offered, such as on-demand rider to vehicle matching for ride-sharing services; (ii) the network supply, such as traffic-responsive tolling or signal control; and (iii) traveler behavior, such as accounting for the heterogeneous preferences of travelers that translate to diverse route choices, mode choices, and departure time choices; as well as heterogeneous driver behavior: from law-abiding risk averse drivers to risk-seeking drivers. 
These traffic simulators are currently being used to tackle a variety of use cases such as the design of: electric-vehicle charging stations, congestion pricing and tolling \citep{Osorio21}, signal control \citep{Osorio15a}, and public transportation expansions.  

However, the most difficult open problem remains that of calibrating both the demand and the supply inputs of these simulators so that they are capable of replicating prevailing, and intricate, traffic patterns of congested metropolitan scale networks. 
More specifically, the most challenging, yet also most important, calibration problem is that of travel demand calibration, traditionally formulated as origin-destination (OD) demand calibration. 

\input{2_litt}
\input{3_method}

\input{4_numerical}
\section{Conclusions}
\label{sec:concl}

This work proposes a metamodel algorithm approach for the OD demand calibration of stochastic high-resolution traffic simulators. The calibration is based on abundant path travel time data. The approach is applied to 6 cities and 9 hours for a total of 54 calibration scenarios. Compared to a commonly used benchmark, SPSA, the proposed approach has enhanced calibration quality, sample efficiency, and robustness to initial points. It improves the fit to field data, as measured by ETA nRMSE, reaching up to a 80.0\% improvement with an average of 43.5\%. In this work, the various hourly calibration problems are solved independently. Future work will extend the proposed algorithm to simultaneously tackle the problem across multiple time intervals. Of interest is also the use of calibrations based on the use of higher-order (i.e., beyond first-order) moments of path ETA statistics. 

\bibliographystyle{IEEEtranN}
\bibliography{biblio_spacetime_co, old_biblio_spacetime}

\end{document}

%% file: 2_litt.tex
\label{sec:litt}

Table~\ref{tab:lit_summary} summarizes a small sample of past OD calibration work. In the literature, the majority of past work has considered the use of conventional segment data \citep{Osorio19b}, such as count, speed, and occupancy. The use of path-level data, such as path travel time (also referred to as estimated time of arrival or ETA) data, has been less explored. Segment data, obtained typically from underground loop detectors, is often spatially sparse, contributing to a higher level of underdetermination for the calibration problem. 
Most past work has considered small networks, especially small highway networks. This is the first work to consider highway networks for entire metropolitan areas, let alone doing so for various cities and across time periods of varying levels of congestion.


\begin{table*}[h!]
    \centering
    \caption{Literature on demand calibration}
    \begin{tabular}{c|l|l|l|l} 
        \textbf{Article} & \textbf{Ground truth data (quantity)} & \textbf{Network size} & \begin{tabular}{@{}l@{}}\textbf{Demand} \\\textbf{dimension * time}\end{tabular} & \textbf{Algorithm(s)} \\ \hline \hline
        Daguano et al. 2023 \citep{Daguano23} & Synthetic segment counts and ETAs & 4 urban intersections & & \begin{tabular}{@{}l@{}}Artificial \\neural networks\end{tabular} \\ \hline
        Vahidi and Shafahi 2023 \citep{Vahidi23} & \begin{tabular}{@{}l@{}}Synthetic subpath ETAs, \\segment counts\end{tabular} & 1,101 urban segments & 462 * 8 & Analytical, LSQR \\ \hline
        Kumarage et al. 2023 \citep{Kumarage23} & Synthetic segment counts (395) & 5,799 urban segments & 3,249 & \begin{tabular}{@{}l@{}}Bi-level hybrid \\(macro, micro)\end{tabular} \\ \hline
        Sha et al. 2023 \citep{Sha23} & \begin{tabular}{@{}l@{}}Real-world segment counts, ETAs,\\and traffic conflicts\end{tabular} & 1 urban corridor & & SPSA \\ \hline
        Ho et al. 2023 \citep{Ho23} & Synthetic segment counts (36) & 1,106 urban segments & 1,615 & \begin{tabular}{@{}l@{}}Metamodel-based \\SPSA\end{tabular} \\ \hline
        Huo et al. 2023 \citep{Huo23} & Synthetic segment counts (28)  & \begin{tabular}{@{}l@{}}Highway corridor with\\3 ramps\end{tabular} & 30 * 8 & \begin{tabular}{@{}l@{}} Metamodel,\\Bayesian optimization\end{tabular} \\ \hline
        
        Vishnoi et al. 2023 \citep{Vishnoi23} & Synthetic segment speeds (2,282) & 15,254 urban segments & 62 & Metamodel \\ \hline
        
        
        Osorio 2019 \citep{Osorio19b} & Synthetic segment counts (172) & \begin{tabular}{@{}l@{}}1,150 arterial and\\expressway segments\end{tabular} & 4,050 & Metamodel \\ \hline
        
        Carrese et al. 2017 \citep{Carrese17} & \begin{tabular}{@{}l@{}}Synthetic path ETAs (12),\\segment counts (32)\end{tabular} & 812 urban segments & & \begin{tabular}{@{}l@{}} Kalman filter (KF):\\LETKF\end{tabular} \\ \hline

        Zhang and Osorio 2017 \citep{Zhang17} & Real-world segment counts (346) & 24,335 urban segments & 2,585 & Metamodel \\ \hline
        Cipriani et al. 2014 \citep{Cipriani14} & \begin{tabular}{@{}l@{}}Synthetic path ETAs (1),\\segment speeds (5), and counts (5)\end{tabular} & 8 urban segments &  & SPSA \\ \hline
        Zhou et al. 2012 \citep{Zhou13} & \begin{tabular}{@{}l@{}}Real-world segment ETAs,\\segment counts, segment occupancies\end{tabular} & \begin{tabular}{@{}l@{}}Highway corridor with\\3 on-ramps/1 off-ramp\end{tabular} &  & Gradient descent \\ \hline
        Barcel\'{o} et al. 2012 \citep{Barcelo12} & Synthetic path ETAs, segment counts & 232 urban segments & 85 * 4 & KF \\ \hline
        
        Ben-Akiva et al. 2012 \citep{Benakiva12} & \begin{tabular}{@{}l@{}}Real-world segment ETAs (52,545),\\segment counts (1,694)\end{tabular} & 3,180 urban segments & 
        2,927 * 16
        & SPSA \\ \hline
        This paper & Real-world path ETAs (1,676) & 18,650 highway segments & 1,676 * 9 & Metamodel \\
        
        
    \end{tabular}
    \label{tab:lit_summary}
\end{table*}

The OD calibration problem, formulated in Section~\ref{sec:method}, involves the calibration of high-resolution (e.g., mesoscopic, microscopic) stochastic simulation-based traffic models. It is a high-dimensional continuous simulation-based optimization problem. A survey of algorithmic approaches is  provided in \citep{Zhang18}. 
Commonly used algorithms include metamodel algorithms \citep{Huo23,Vishnoi23,Osorio19a, Osorio19b, Zhang17}, 
simultaneous perturbation stochastic approximation (SPSA, e.g., \citep{Cipriani14, Benakiva12}), genetic algorithms (GA, e.g., \citep{Chiappone16}), and Kalman filtering (KF, e.g., \citep{Carrese17, Barcelo12}). 

This paper focuses on the metamodel approach. First proposed in \citep{Osorio13}, it stands out as a family of compute- and sample-efficient techniques. It achieves this by combining a physics-based approximation of the loss function with a general-purpose statistical model. The infusion of simplified physics information allows the algorithm to identify good quality solutions with few simulation calls (i.e., it is sample efficient). This also enables it to become robust to the quality of initial points. Compute-efficiency and the ability to tackle high-dimensional instances are achieved by tackling the simulation-based optimization problem through solving a sequence of analytical and differentiable problems, which can be solved efficiently with traditional gradient-based solvers.  
In this paper, we extend past algorithmic formulations to allow it's use for path-level, rather than segment-level, field data. In particular, we consider its use in calibrating demand based on path ETA data.

The contributions of this paper are as follows. 
\begin{itemize}
    \item This paper is the first to showcase the ability to perform OD calibration for 6 major US cities. For each city, we calibrate demand for 9 time intervals, leading to a total of 54 calibration scenarios. This differs from past work that considers a single city or location, hindering generalizability of the findings.
    \item For each city, we illustrate the ability of the proposed approach to calibrate all ranges of congestion: from free-flowing to highly congested conditions. We do so by calibrating demand for various hours of a day. This differs from past work that considers limited congestion scenarios.
    \item For each city, we calibrate the entire metropolitan-scale highway network. This contrasts with past work that focuses on highway corridors with limited ramps (e.g., \citep{Huo23,Zhou13}) or network combining arterials and expressways (e.g., \citep{Osorio19b}). 
    \item The proposed algorithm builds upon an existing metamodel optimization framework \citep{Osorio13}. We extend the framework to utilize high-resolution path ETAs as field data. By incorporating path ETAs, the framework gains access to route-level insights, transcending the limitations of conventional segment-specific sensor counts. The use of path ETAs, obtainable from readily accessible and cost-effective sources without requiring expensive sensor infrastructure, significantly contributes to the framework's scalability.
    \item Through extensive numerical experiments, the proposed approach is shown to be robust to initial points and to  congestion levels. Moreover, it is sample efficient: it identifies points with improved fit to field data with few simulation function evaluations. This is important for cities that often tackle calibration problems with a tight compute time budget. The approach outperforms the well-established SPSA benchmark algorithm: it improves the fit to field data by an average of 43.5\%, and in some cases, the improvement reaches as high as 80.0\%.
\end{itemize}



Section~\ref{sec:method} formulates the calibration problem and presents the algorithmic approach. Numerical case studies  are presented in Section~\ref{sec:numerical}, followed by conclusions (Section~\ref{sec:concl}). 


%% file: 3_method.tex
\section{Methodology}
\label{sec:method}

\subsection{Calibration Problem Formulation}

The origin-destination (OD) demand calibration problem aims to calibrate, for a given time interval, the expected number of trips from a given origin zone to a given destination zone. We consider here what is known as static OD calibration where a demand is calibrated for a single time interval. This differs from dynamic OD calibration where demand is calibrated simultaneously for multiple time intervals. We also consider offline calibration problems which differ from online problems that are solved in real-time.

The origin and destination zones are typically defined based on spatial partitions of the road network that are commonly used by transportation planners and engineers, such as census tracts.  
The vector of OD demands is often represented as a matrix, and hence is also referred to as an OD matrix. 

To formulate the calibration problem, we introduce the following notation:

\begin{tabular}{l}
\textbf{Problem variables:}\\
$x$: OD demand vector,\\
$f(x)$: simulation-based  loss (or objective) function,\\
$u_1$: vector of endogenous simulation variables (e.g., \\
\hspace{17pt}segment queue-lengths),\\
$E[Y_p(x,u_1;u_2)]$: expected (simulation-based) travel\\
\hspace{17pt}time for path $p$,\\

\textbf{Problem parameters:}\\
$\mathcal{P}$: set of paths with ground truth (GT) travel times,\\
$|\mathcal{P}|$: cardinality of set $\mathcal{P}$,\\
$u_2$: vector of exogenous simulation parameters (e.g.,\\
\hspace{17pt}road network topology, segment attributes such\\
\hspace{17pt}as length, number of lanes, and free-flow speed),\\
$y_p^{\text{GT}}$: GT measurement of average travel time for path\\
\hspace{17pt}$p$,\\
$x_U$: OD demand upper bound vector.\\
\end{tabular} \\

The problems is formulated as a continuous simulation-based optimization (SO) problem as follows.
\begin{align}
\min_{x} \quad & f(x)=\frac{1}{|\mathcal{P}|}\sum_{p \in \mathcal{P}}(y_p^{\text{GT}} - E[Y_p(x,u_1;u_2)])^2
\label{eq:f} \\
\textrm{s.t.} \quad & 0 \le x \le x_U.
\label{eq:bounds}
\end{align}

The main goal of the demand calibration problem is to identify an OD demand vector that once simulated replicates traffic statistics from field measurements. The field measurements are referred to as ground truth (GT) data or GT metrics. Here we aim to minimize the distance between field measured path travel times and the corresponding expected simulation-based path travel times. 
Problem formulations differ in their choice of the distance metric which is used to measure the discrepancy between GT statistics and their corresponding simulated statistics. We use the most common formulation, the weighted least squares. 

The OD calibration problem is underdetermined, meaning that there are a continuum of OD vectors that provide a statistically equivalent fit to the GT data. To address this, the loss function (Eq.~\eqref{eq:f}) often includes a regularization term which  serves to bias the solution of the problem towards a specific OD vector, which is most often an OD calibrated years back from past census or past simulation studies. Since this work focuses on the ability of performing calibrations for various and arbitrary cities, we do not assume availability of such ODs and therefore do  not consider a regularization term. 

The main challenges of Problem~\eqref{eq:f}-\eqref{eq:bounds} are as follows. Firstly, it is a high-dimensional SO problem. SO problems with dimension in the low-hundreds and up are considered high-dimensional. In our case, the dimension of $x$ ranges from hundreds to thousands. Secondly, the estimation of $f$  (Eq.~\eqref{eq:f}), and more specifically of  
$E[Y_p(\cdot)]$, is computationally costly. A single function evaluation (i.e., one simulation run) of a metropolitan-scale network can range from dozens of minutes to several hours. Hence, there is a need to consider sample-efficient algorithms that can identify points with good solutions while requiring few function evaluations. 
Thirdly, the considered simulators are stochastic and their loss function is  non-differentiable. Hence, traditional gradient-based algorithms that assume deterministic loss functions are not appropriate. Moreover, in high-dimensional setting, gradient-based algorithms that account for stochasticity typically encounter challenges due to the high variance of the gradient estimator. 
Given these challenges, we opt for metamodel algorithms that have been proven to scale to high-dimensions while remaining sample-efficient. Moreover, they do not require the evaluation of simulation-based gradients, making them  robust to simulator stochasticity. 

\subsection{Methodology}

The main idea of the metamodel framework is to  formulate a surrogate (or metamodel) of the simulation-based objective function, and to solve Problem~\eqref{eq:f}-\eqref{eq:bounds} by solving a series of approximate optimization problems that can be solved in a computationally efficient way and without the need for large samples from the costly simulator. The key to tackling high-dimensional problems while remaining sample efficient has been to: (i) define surrogate models that are deterministic and differentiable, such that traditional (and  widely used and studied) deterministic gradient-based algorithms can be used to tackle the approximate problem; and (ii) define surrogates that infuse (approximate) physics information  from Problem~\eqref{eq:f}-\eqref{eq:bounds} so that OD vectors with good simulated performance can be identified  with  few simulator function evaluations. 

For a detailed presentation of the general metamodel optimization framework we refer the reader to \citep{Osorio19c}. At epoch (or iteration) $k$, of the proposed SO algorithm, we tackle   Problem~\eqref{eq:f}-\eqref{eq:bounds} by solving the below non-simulation based, analytical, and differentiable optimization problem based on the following notation. \\


\begin{itemize}
    \item[] $k$: SO algorithm epoch,
    \item[] $m_k$: metamodel function at epoch $k$,
    \item[] $\phi$: functional, i.e., general-purpose, component of the metamodel $m_k$,
    \item[] $f_A(x)$: physics-based component of the metamodel,
    \item[] $\beta_k$: parameter vector of metamodel $m_k$,
    \item[] $\beta_{k,j}$: scalar element $j$ of parameter vector $\beta_k$,
    \item[] $x_z$: element $z$ of OD vector $x$,
\end{itemize}

\textbf{Variables of the analytical network model:}
\begin{itemize}
    \item[] $y_p^\text{A}$: analytical approximation of travel time of path $p$,
    \item[] $\lambda$: vector of demand for all segments in the network, element $i$ is denoted $\lambda_i$,
    \item[] $k_i$: density of segment $i$,
    \item[] $v_i$: (space-mean) speed of segment $i$,
\end{itemize}

\textbf{Problem parameters:}
\begin{itemize}
    \item[] $\ell_i$: length  of  segment $i$,
    \item[] $n_i$: number of lanes of segment $i$,
    \item[] $v^{\text{max}}_i$: maximum speed (i.e., speed-limit) of segment $i$,
    \item[] $A$: matrix that maps OD demand to segment demand,
    \item[] $\alpha_1, \alpha_2$: segment fundamental diagram parameters,
    \item[] $\kappa_1$: density scaling factor,
    \item[] $k^{\text{jam}}$: jam density,
    \item[] $v^{\text{min}}$: minimum segment speed,
    \item[] $\mathcal{I}$: set of all segments in the network,
    \item[] $\mathcal{I}_p$: set of segments of path $p$.
\end{itemize}

\begin{align}
\min_{x} \quad & m_k(x;\beta_k)=\beta_{k,0}f_A(x) + \phi(x;
\beta_k)
\label{eq:metam1}\\
\textrm{s.t.} \quad & f_A(x) = 
\frac{1}{|\mathcal{P}|}\sum_{p \in \mathcal{P}}(y_p^{\text{GT}} - y_p^\text{A})^2
\label{eq:metam2}\\
\quad & \phi(x;\beta_k)=\beta_{k,1}+\sum_{z\in\mathcal{Z}}{\beta_{k,z+1}x_z}
\label{eq:metam3}\\
\quad & \lambda = Ax
\label{eq:metam4}\\
\quad & k_i = \frac{\kappa_1 k^\text{jam} }{n_i}\lambda_i, \ \forall i \in \mathcal{I}
\label{eq:metam5}\\
\quad & v_i = v^{\text{min}} + (v^{\text{max}}_i - v^{\text{min}})\left(1 - \left(\frac{k_i}{k^{\text{jam}}}\right)^{\alpha_1}\right)^{\alpha_2}, \  \forall i \in \mathcal{I}
\label{eq:metam6}\\
\quad & y_p^\text{A} = \sum_{i \in \mathcal{I}_p} \frac{\ell_i}{v_i}, \ \forall p \in \mathcal{P}
\label{eq:metam7}\\
\quad & 0 \le x \le x_U.
\label{eq:metam8}
\end{align}

In a nutshell, Problem~\eqref{eq:metam1}-\eqref{eq:metam8} uses a simplified, macroscopic analytical network model, defined by Eqs.~\eqref{eq:metam4}-\eqref{eq:metam7}, to approximate the simulation-based metrics of interest: expected path travel times. In particular, the fundamental diagram that models the speed-density relationship as in Eq.~\eqref{eq:metam6} is based on \citep{May67}. The above system of equations uses $y_p^\text{A}$ (Eq.~\eqref{eq:metam7}) as an approximation of $E[Y_p(\cdot)]$ in Eq.~\eqref{eq:f}. The simplified network model's loss function approximation, represented by Eq.~\eqref{eq:metam2}, serves as an approximation of the simulation-based loss function, which is given by Eq.~\eqref{eq:f}. Moreover, the augmentation with the polynomial term $\phi(\cdot)$ makes a more general surrogate model $m_k(\cdot)$ (Eq.~\eqref{eq:metam1}) which approximates with greater fidelity the simulation-based loss function. 

For the first time, we derive and use analytical approximations of path metrics, in this case path travel time metrics. Past work had considered the use of either segment count or segment speed metrics. The analytical network model is also simplified relative to past work. In particular, for all segments, we consider: (i) common exponent parameters of the fundamental diagrams ($\alpha_1$ and $\alpha_2$ of Eq.~\eqref{eq:metam6}); (ii) common minimum speed parameters ($v^\text{min}$ of  Eq.~\eqref{eq:metam6}); and (iii) common jam density parameters  ($k^\text{jam}$ of Eqs.~\eqref{eq:metam5}-\eqref{eq:metam6}). Unlike past work that considered the use of segment speed data \citep{Vishnoi23}, a different functional form is used for the  fundamental diagram. This simplified model assumes that route choice is exogenous and fixed, just as in \citep{Osorio19a} yet unlike the formulation \citep{Osorio19b} that allows for endogenous route choice.

%% file: 4_numerical.tex
\section{Multi-city metropolitan-scale case studies}
\label{sec:numerical}
\subsection{Experimental Design}
\begin{figure}[htb]
	\center
	\includegraphics[width=0.20\textwidth]{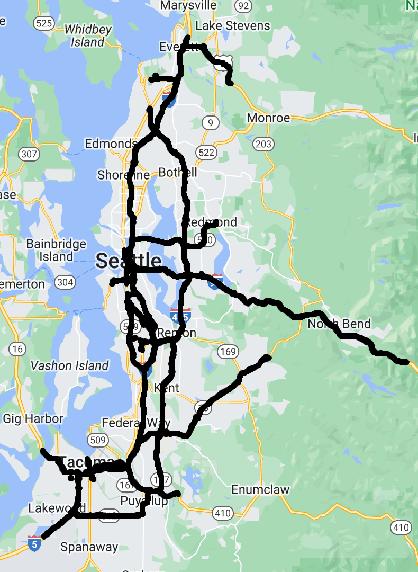}
	\caption{Highway network of the Seattle metropolitan area.}
    \label{fig:seattle_highway_network}
\end{figure}

We evaluate our algorithm's performance on calibrating OD demand for 6 real-world metropolitan highway networks: Seattle, Denver, Philadelphia, Boston, Orlando, and Salt Lake City. The scope of our study encompasses a highway system that excludes signalized arterial roads. Highway ODs are ramp-to-ramp trips that utilize only highway and ramp road segments. We use SUMO simulation software \citep{Sumo18} to model the highway networks. The highway network of Seattle metropolitan area is illustrated in Figure \ref{fig:seattle_highway_network}. For each city, we calibrate hourly ODs from 2pm to 11pm of a typical weekday. This period includes the afternoon peak hours as well as the off-peak hours, allowing us to evaluate the algorithm's performance under different congestion levels. For each scenario (i.e., city and hour combination), we use aggregated and anonymized path ETA data for each OD pair. 
Characteristics of the considered networks are given in  Table~\ref{tab:network_spec}. The table lists the number of segments in the network, and gives the maximum, across the various hours, of the number of ODs. This also represents the maximum dimensionality encountered in the series of OD calibration problems. We benchmark our approach versus the commonly used SPSA algorithm \citep{Spall03}.
For each scenario and each approach (metamodel or SPSA), we initialize the approach with a common and feasible random OD. 

\begin{table}[h!]
    \centering
    \caption{Scenario network specification}
    \begin{tabular}{c|c|c} 
        \textbf{Metro area} & \textbf{Number of segments} & \begin{tabular}{@{}c@{}}\textbf{Max hourly}\\\textbf{OD dimensionality}\end{tabular} \\ \hline \hline
        Seattle & 18,650 &  1,676 \\ \hline
        Denver & 15,647 & 1,203 \\ \hline
        Philadelphia & 10,709 & 890 \\ \hline
        Boston & 6,689 & 690 \\ \hline
        Orlando & 8,028 & 636 \\ \hline
        Salt Lake City & 11,487 & 617 \\
    \end{tabular}
    \label{tab:network_spec}
\end{table}


\begin{figure*}[htb]
	\center
	\includegraphics[width=1.75\columnwidth]{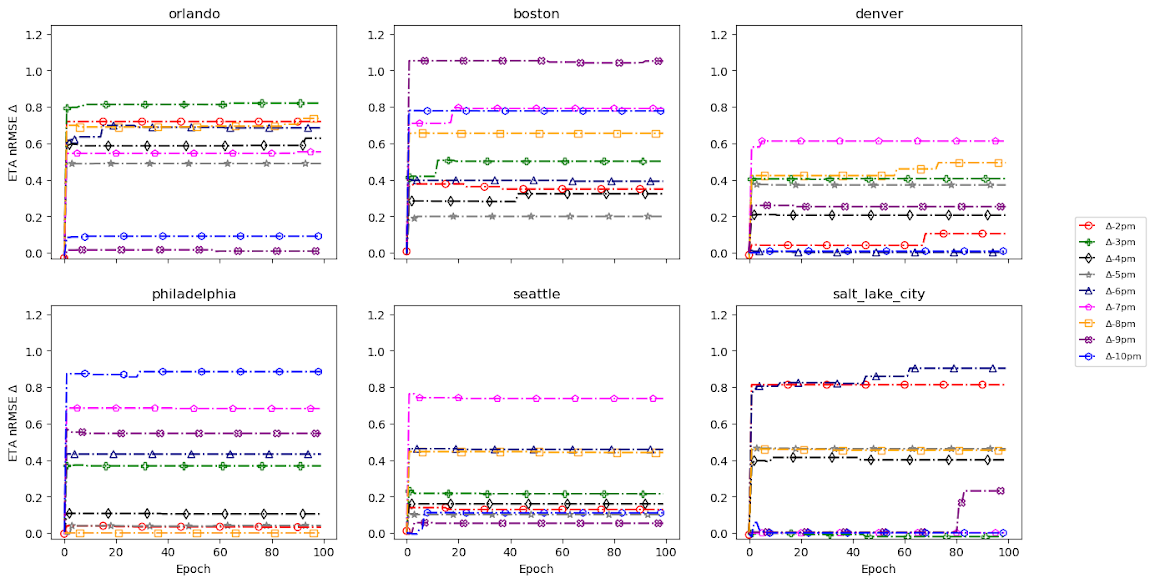}
    \caption{Difference $\Delta$ of metamodel relative to SPSA in terms of ETA nRMSE as a function of epochs.}
    \label{fig:eta_nrmse_diff}
\end{figure*}

\subsection{Results and Discussion}



The calibration quality (both metamodel and SPSA) is assessed in terms of the normalized root mean square error (nRMSE) of path ETAs as a function of algorithmic epochs, where nRMSE is defined by Eq. \eqref{eq:nRMSE}:

\begin{align}
nRMSE = \frac{|\mathcal{P}|}{\sum_{p \in \mathcal{P}} y_p^{GT}} \sqrt{\frac{1}{|\mathcal{P}|}\sum_{p \in \mathcal{P}} \left(E[Y_p(x, u_1;u_2)] - y_p^{GT}\right)^2}.
\label{eq:nRMSE}
\end{align}





Figure~\ref{fig:eta_nrmse_diff} provides a visual representation of the metamodel's improved performance compared to SPSA. This is accomplished by plotting the difference $\Delta$ in ETA nRMSE between the metamodel and SPSA at each epoch for each hour, specifically subtracting the ETA nRMSE of metamodel from that of the SPSA. Each of the 9 hours are represented by a separate marker and color. With each plot dedicated to a single city, it becomes evident that the metamodel consistently outperforms SPSA, rapidly establishing (even in the first few epochs) and steadily increasing its advantage within the computational resources available. The maximum ETA nRMSE difference $\Delta$ is 1.1 (Boston), with an average of 0.4 across all cities and hours. This sustained improvement ultimately leads to the metamodel generating higher quality calibrated demand than SPSA, consistently across various cities and congestion levels.

In particular, the most substantial improvement in calibration performance within a single hour occurs during the afternoon peak period (i.e., 6-7pm), which experiences greater traffic congestion. The highest single-hour calibration performance improvement (i.e., ETA nRMSE difference $\Delta$ normalized by the ETA nRMSE of SPSA) reaches 80.0\% at 6PM for Salt Lake City, where the average improvement across all hours is 49.7\%. In Boston, the peak performance improvement is 76.8\% at 7PM, with an average improvement of 54.2\%. Regarding all 54 calibration scenarios (i.e., 6 cities with 9 hours each), the average improvement is 43.5\%.

While the empirical analysis demonstrates a substantial enhancement in calibration quality compared to the baseline SPSA method, achieving satisfactory performance across all metropolitan areas and time frames remains challenging. Several avenues for future refinement exist to minimize this error discrepancy: (1) network refinement: enhance data accuracy of segment attributes, such as lane-to-lane connectivity, speed limit, number of lanes; (2) various traffic data integration: leverage supplementary data sources beyond path ETAs to compensate for inherent limitations and improve overall model fidelity.
These enhancements are anticipated to further narrow the gap between simulated and observed traffic dynamics, leading to more robust and reliable calibration outcomes.



